\documentclass[aps,english,prb,amsmath,amsfonts,amssymb,superscriptaddress,twocolumn,showpacs,citeautoscript]{revtex4-1}
\usepackage[T1]{fontenc}
\usepackage[latin9]{inputenc}
\usepackage{amsmath}
\usepackage{amssymb}
\usepackage{wasysym}
\usepackage{esint}
\usepackage{graphicx}
\usepackage{times}
\usepackage{nicefrac}
\usepackage{appendix}
\usepackage{textcase}
\usepackage{subfigure}
\usepackage{empheq}
\usepackage{color}
\usepackage{leftidx}
\usepackage{makecell}
\usepackage{stackrel}

\makeatletter
\@ifundefined{textcolor}{}
{%
 \definecolor{BLACK}{gray}{0}
 \definecolor{WHITE}{gray}{1}
 \definecolor{RED}{rgb}{1,0,0}
 \definecolor{GREEN}{rgb}{0,1,0}
 \definecolor{BLUE}{rgb}{0,0,1}
 \definecolor{CYAN}{cmyk}{1,0,0,0}
 \definecolor{MAGENTA}{cmyk}{0,1,0,0}
 \definecolor{YELLOW}{cmyk}{0,0,1,0}
}

\renewcommand{\vec}[1]{\mathbf{#1}}
\renewcommand{\Re}{\operatorname{Re}}
\renewcommand{\Im}{\operatorname{Im}}

\newcommand{\sym}{\operatorname{sym}}
\newcommand{\asym}{\operatorname{asym}}

\newcommand{\s}{\sigma}
\renewcommand{\b}{\beta}

\newcommand{\GF}{G^\mathrm{F}}

\newcommand{\add}[1]{\if\a\b{{\color{red} #1}}\else{#1}\fi}

\newcommand{\citeasnoun}[1]{Ref.~\onlinecite{#1}}

\newcommand{\citesasnoun}[1]{Refs.~\onlinecite{#1}}

\renewcommand{\eqref}[1]{Eq.~\ref{eq:#1}}

\newcommand{\Eqref}[1]{Equation~\ref{eq:#1}}
\newcommand{\figref}[1]{Fig.~\ref{fig:#1}}
\newcommand{\Figref}[1]{Figure~\ref{fig:#1}}

\newcommand{\secref}[1]{Sec.~\ref{sec:#1}}

\def\ro{\vec{x}}
\def\rs{\vec{y}}
\newcommand{\mat}[1]{{#1}}
\newcommand{\matvec}[1]{{#1}}

\newcommand{\trace}[1]{{\rm Tr} \left[ #1 \right]}

\newcommand{\vb}{\mathbf}

\newcommand{\Tr}{\text{Tr }}

\makeatother

\usepackage{babel}
\begin{document}

\title{Fluctuating volume--current formulation of electromagnetic
  fluctuations in inhomogeneous media: incandecence and luminescence
  in arbitrary geometries}


\author{Athanasios G. Polimeridis}
\affiliation{Skolkovo Institute of Science and Technology, Moscow, Russia}
\author{M. T. H. Reid}
\affiliation{Department of Mathematics, Massachusetts Institute of Technology, Cambridge, MA 02139, USA}
\author{Weiliang Jin}
\affiliation{Department of Electrical Engineering, Princeton University, Princeton, NJ 08544, USA}
\author{Steven G. Johnson}
\affiliation{Department of Mathematics, Massachusetts Institute of Technology, Cambridge, MA 02139, USA}
\author{Jacob K. White}
\affiliation{Department of Electrical Engineering and Computer Science, Massachusetts Institute of Technology, Cambridge, MA 02139, USA}
\author{Alejandro W. Rodriguez}
\affiliation{Department of Electrical Engineering, Princeton University, Princeton, NJ 08544, USA}

\begin{abstract}
  We describe a fluctuating volume--current formulation of
  electromagnetic fluctuations that extends our recent work on heat
  exchange and Casimir interactions between arbitrarily shaped
  homogeneous bodies [Phys. Rev. B. 88, 054305] to situations
  involving incandescence and luminescence problems, including thermal
  radiation, heat transfer, Casimir forces, spontaneous emission,
  fluorescence, and Raman scattering, in inhomogeneous media. Unlike
  previous scattering formulations based on field and/or surface
  unknowns, our work exploits powerful techniques from the
  volume--integral equation (VIE) method, in which electromagnetic
  scattering is described in terms of volumetric, current unknowns
  throughout the bodies. The resulting trace formulas (boxed
  equations) involve products of well-studied VIE matrices and
  describe power and momentum transfer between objects with spatially
  varying material properties and fluctuation characteristics. We
  demonstrate that thanks to the low-rank properties of the associated
  matrices, these formulas are susceptible to fast-trace computations
  based on iterative methods, making practical calculations
  tractable. We apply our techniques to study thermal radiation, heat
  transfer, and fluorescence in complicated geometries, checking our
  method against established techniques best suited for homogeneous
  bodies as well as applying it to obtain predictions of radiation
  from complex bodies with spatially varying permittivities and/or
  temperature profiles.
\end{abstract}

\maketitle

\section{Introduction}



Quantum and thermal fluctuations of charges give rise to a wide range
of electromagnetic phenomena; these include luminescence from active
media, e.g. fluorescence and spontaneous
emission~\cite{Siegman89,Agarwal74,LeRu08}, the finite linewidth of
lasers near threshold~\cite{Gordon55,Matloob97}, thermal radiation and
heat transfer from hot
objects~\cite{PolderVanHove71,Joulain05,Chen05,Carey06,Fu06,Volokitin07,
  Zhang07,BasuZhang09,Otey14:review}, and dispersive interactions
(Casimir forces) between nearby
surfaces~\cite{Casimir48:polder,casimir,Buhmann07,Genet08,Bordag09:book,Rodriguez11:review,Rodriguez14:review}. Fluctuation-driven
effects are not only responsible for many naturally occurring processes
but are also poised to take an increasingly active role in emerging
nanotechnologies~\cite{Zhang07,BasuZhang09}, spurring interest in the
study and engineering of complex shapes that could dramatically alter
their behavior~\cite{Otey14:review,Rodriguez14:review}. Although
rooted in similar principles, the physical mechanisms behind each of
these processes vary considerably, leading to theoretical descriptions
that differ both in their formulation and implementation. Ultimately,
however, all such calculations reduce to a series of classical
scattering problems~\cite{Kahnert03,Johnson11:review} that until
recently remained largely specialized to situations involving simple,
high--symmetry geometries, e.g. planar and spherical objects.

In this manuscript, we present a framework for the general-purpose
calculation of many different incandescence and luminescence
processes, including fluorescence, spontaneous emission, thermal
radiation, heat transfer, and Casimir forces in arbitrary
geometries. In particular, we derive a fluctuating volume--current
(FVC) formulation of electromagnetic fluctuations that exploits
techniques from the volume--integral equation (VIE) formulation of
electromagnetic scattering~\cite{Chew_book,Polimeridis2014} and which
expands the range and validity of current methods to situations
involving inhomogeneous media. Although FVC is similar in spirit to
our previous fluctuating surface--current (FSC)
methods~\cite{RodriguezReid12:FSC,RodriguezReid12:long}, unlike FSC
our new approach is not limited to piecewise-homogeneous
objects. Here, the unknowns are volume currents within objects rather
than surface currents as in FSC, and can therefore easily handle more
complex structures, including inhomogeneous bodies with temperature
gradients or spatially varying permittivities. In contrast to recently
developed scattering-matrix
methods~\cite{Lambrecht06,Milton08:scat,Rahi09:PRD,bimonte09,Biehs11:apl,Guerout12,Messina11,Kruger11,OteyFan11,Golyk12,Kruger12,Marachevsky12,Lussange12},
the FVC and FSC methods do not require a separate basis of
incoming/outgoing wave solutions to be selected (a potentially
difficult task in geometries involving interleaved objects or complex
structures favoring nonuniform spatial resolution), although VIE can
be used to compute the scattering matrix if desired. We show that
regardless of which quantity is computed, the final expressions for
power and momentum transfer are based on simple trace formulas
involving well-studied VIE and current--current correlation matrices
that encode the spectral properties of fluctuating sources. We find
that while the number of VIE unknowns is large compared to scattering
or FSC formulations, the associated VIE matrices admit low-rank
approximations that turn out to significantly reduce the complexity of
trace evaluations, making practical calculations tractable. We
validate the FVC method by checking its predictions against known
solutions for homogeneous objects and then apply it to calculate
thermal radiation, heat transfer, and fluorescence from compact
objects (spheres, ellipsoids, and cubes) with spatially varying
permittivities and temperature gradients. The same trace formulas can
be readily adapted to obtain the angular distribution of far-field
radiation, which we illustrate by providing new predictions of
directional emission from inhomogeneous objects. 
the
As explained below, while VIE methods can be applied to
arbitrary geometries, they are particularly advantageous in situations
where object sizes are on the order of (or smaller) than the relevant
wavelengths, providing a useful complement to well-established
techniques better suited for the study of arbitrary geometries with
lengthscales that are large or small compared to the relevant
electromagnetic wavelengths, e.g. proximity
approximations~\cite{Bordag09:book,Sasihithlu11}.


Electromagnetic fluctuation phenomena can be roughly divided into two
categories: incandescence and luminescence
problems. \emph{Incandescence} refers to electromagnetic radiation
from objects generated by the quantum and thermal motion of charged
particles in matter, whereas \emph{luminescence} refers to incoherent
emission of light from non-thermal sources. The oldest and most
well-studied manifestation of incandescence is the familiar glow of
objects---thermal radiation---that occurs when an object is heated
above the temperature of its surrounding
environment~\cite{Reif:stat,Landau:stat}. Although Planck's law was
not more than a century ago at the center of vigorous controversy
which helped establish the foundations of quantum
mechanics~\cite{Planck1901}, much of our recent interest in this
phenomenon spawns from its profound impact on energy and related
nanotechnologies. Interest in complex designs is also fueled by our
increasing ability to engineer selective and even dynamically tunable
emitters and detectors at wavelengths for which there is currently a
lack of coherent
sources~\cite{Greffet01,Laroche06:prl,Zhang07,Schuller09,Liu11,Inoue14},
in addition to solar-energy harvesting
applications~\cite{Rephaeli09,Sergeant10,Rinnerbauer12,Gan13,lenert14}. In
addition to radiation, fluctuations can also mediate heat
exchange~\cite{PolderVanHove71,Pendry99,Chen05} and
interactions~\cite{Casimir48:polder,Joulain05,Dalvit11:review,ReidRo12:review,Rodriguez14:review}
(known as Casimir forces) between objects---unlike heat exchange,
Casimir interactions persist even at equilibrium and are known to
arise primarily due to contributions of quantum rather than
finite-temperature fluctuations. One fundamental distinction between
``near-field'' effects (between objects at wavelength-scale
separations or less) and the more familiar ``far-field'' phenomena
(separations $\gg$ wavelength) is that the former can be significantly
enhanced by the contributions of evanescent
waves~\cite{Rytov89,PolderVanHove71,Loomis94,Pendry99}, growing in a
power-law fashion with decreasing object separations.  As a result,
the heat transfer between real materials can exceed the predictions of
the Planck blackbody law by orders of magnitude~\cite{BasuZhang09} and
quantum forces can even reach atmospheric pressures at nanometric
lengthscales~\cite{Rodriguez14:review}, motivating interest in complex
designs that can be tailored for various applications, including
thermophotovoltaic energy
conversion~\cite{Pan00,Laroche06,Messina13,Ilic12}, nanoscale
cooling~\cite{Tschikin12,StGelais14}, and MEMS
design.~\cite{Serry98,hochan1,DelRio05}

Until very recently, however, calculations and experiments remained
focused on planar structures and simple approximations
thereof~\cite{Mulet02,Joulain05,Chen05,Carey06,Fu06,Volokitin07,Zhang07,BasuZhang09,Otey14:review}.
Since all such thermal effects arise due to the presence of
fluctuating current sources, from the perspective of calculations
their descriptions reduce to a series of classical scattering
calculations involving fields due to
currents~\cite{Johnson11:review,Otey14:review}, the spectral
characteristics of which are related to the underlying physical means
of excitations. In the case of incandescence, they are determined by
the thermal and dissipative properties of materials via the well-known
fluctuation--dissipation theorem
(FDT)~\cite{Eckhardt84,Landau:stat2}. Naively, this involves repeated
calculations of electromagnetic Green's functions throughout the
bodies, which can prove prohibitive for complex objects where the
latter must be computed numerically, especially due to the broad
bandwidth associated with thermal fluctuations, but it turns out that
more sophisticated formulations
exist~\cite{Rodriguez14:review,Otey14:review}. These include time- and
frequency-domain methods where the power transfer or force on an
object is obtained via integrals of the flux or Maxwell stress tensor,
or equivalently electromagnetic Green's functions, along some
arbitrary surface enclosing the
body~\cite{Narayanaswamy08:spheres,RodriguezMc09:PRA,McCauleyRo10:PRA,Xiong10,OteyFan11,RodriguezIl11,Narayanaswamy13,Liu13}. Recent
techniques forgo surface integrations altogether in favor of
unfamiliar but more efficient expressions involving traces of either
scattering~\cite{Biehs08,bimonte09,Messina11,McCauleyReid12,Guerout12,Marachevsky12,Kruger12,Golyk12}
or
boundary-element~\cite{RodriguezReid12:FSC,RodriguezReid12:long,RodriguezReid12:nonmon}
matrices. Regardless of the choice of unknowns, in practical
implementations the latter are expanded in terms of either delocalized
spectral bases (e.g. Fourier or Mie series) best suited for
high--symmetry geometries, or geometry-agnostic localized bases
(piecewise polynomial ``element'' functions) defined on meshes or
grids and applicable to arbitrary objects~\cite{Johnson11:review}.
While there has been much progress so far, these methods have yet to
be generalized to handle structures with temperature gradients or
varying permittivities.

Temperature gradients can arise for instance due to the interplay of
phonon and photon
transport~\cite{cahill2002thermometry,cahill2003nanoscale}, such as in
heterogeneous structures with disparate thermal conductivities,
including chalcogenide/metal interfaces~\cite{xiong2009inducing,
  liang2012ultra} or quartz-platium-polymer
structures~\cite{fenwick2009thermochemical}, or in graphene-based
devices~\cite{islam2013role}. Temperature gradients have also been
observed in atomic force
microscopes~\cite{king2013heated,biehs2012nanoscale} and
nanowires~\cite{yeo2014single}, as well as in situations involving
irradiated particles immersed in
fluids~\cite{merabia2009critical,Baffou13,govorov2006gold,fang2013plasmon,hu2013photo,baffou2014deterministic,vu2013gold,jonsson2014nanoplasmon,letfullin2008ultrashort,pustovalov2005theoretical},
magnetic nanocontacts~\cite{petit2012understanding}, or microcavities
subject to strong photothermal
effects~\cite{sun2013nonlinear}. Material inhomogeneities also arise
in microcavity lasers stemming from nonlinear
effects~\cite{Pick15:cmt}. Surprisingly, there are only a handful of
calculations involving non-isothermal particles, including calculation
of radiation from atomic gases in shock-layer structures with linear
temperature gradients~\cite{nelson1971continuum} or calculations of
large-radii spheres based on Mie series or related semi-analytical
expansions~\cite{dombrovsky2000thermal,li2012radiative}. As we show in
a separate publication, temperature gradients in inhomogeneous bodies
can lead to a number of interesting effects, including highly
directional thermal emission~\cite{Jin15:thermal}.

Luminescence, like incandescence, involves incoherent emission of
light due to quantum and thermal fluctuations of charges, but differs
in that excitations are driven by coherent rather than thermal
sources. Examples include spontaneous emission, Raman scattering, and
fluorescence from active media externally pumped by coherent
light~\cite{le2005surface,LeRu08,kneipp2006surface}. Although the
spectral properties of fluctuating currents depend on complicated and
often nonlinear light--matter interactions, the resulting radiation is
incoherent and can be modeled by exploiting scattering techniques
similar to those employed in incandescence
problems~\cite{kneipp2006surface}. There are however many important
differences between these two classes of problems.  For instance, the
luminescence spectrum of many emitters is relatively narrow (involving
wavelengths close to material resonances) and this has implications
for calculations which favor frequency as opposed to time-domain
techniques (the latter being better suited for broad-bandwidth
processes). Furthermore, while many thermal radiation problems involve
objects with uniform temperature distributions, the properties of
current fluctuations excited by external pumps depend sensitively on
the inputs and can change dramatically and \emph{continuously}
throughout the bodies, which is problematic for SIE/FSC formulations
based on piecewise homogeneity. Such a situation arises for instance
in the fluorescence from objects with features $\sim$ incident
wavelengths, where resonant absorption can lead to significant spatial
variations in the amplitudes of the fluctuating
currents~\cite{LeRu08}.

Until recently, the fluorescence or Raman emission pattern of small
particles was obtained by analytical methods based on Mie series or
related basis
expansions~\cite{chew1976model,druger1984radiation}. More recent
techniques for studying luminescence from arbitrarily shaped particles
instead rely on numerical
techniques~\cite{myroshnychenko2008modelling}, most commonly
time-domain
methods~\cite{li2007fdtd,yang2010fdtd,rogobete2007design,mohammadi2008gold,mohammadi2010fluorescence,musa2013computational},
and include studies of bowtie antennas~\cite{kinkhabwala2009large},
nanostars~\cite{hao2007plasmon}, conical
tips~\cite{richards2003tip,bian1995single,cade2007plasmonic},
dimers~\cite{dhawan2009comparison}, and thin
films~\cite{yi2015experimental}. Frequency domain methods include
finite-element~\cite{micic2003finite,kottmann2001dramatic},
boundary-element~\cite{teperik2011numerical}, and discrete dipole
approximation
(DDA)~\cite{hao2004electromagnetic,zou2005silver,hao2004optical,edalatpour2015convergence}
methods. These tools have been exploited for instance to demonstrate
that both shape and material degrees of freedom can be used to tailor
particle emission, making it possible to enhance fluorescence and
Raman processes~\cite{le2005surface,LeRu08,kneipp2006surface} as well
as obtain unusual angular emission
patterns~\cite{hill2000enhanced,janssen2010efficient,schick2014multifunctional};
even more recently, there has been interest in studying effects
related to active (non-Hermitian)
systems~\cite{yoo2011quantum,miri2014scattering,hodaei2014pt,chong2011p}. In
most cases (with a few exceptions~\cite{myroshnychenko2008modelling}),
the total radiated power in a given direction is computed by directly
summing the contribution of individual emitters inside the objects,
requiring repeated evaluation of Green's functions over both volumes
and surfaces. In addition, many calculations rely on approximations in
which the effect of the incident drive is either approximated or
entirely neglected~\cite{hill2001fluorescence} or where only the
radiation from a partial set of emitters inside the objects is
obtained~\cite{Agostino13}. Our FVC--VIE approach not only removes
limitations associated with such approximations by fully accounting
for both the emission and excitation-dependent properties of all
fluctuating sources, but introduces new trace-formulas that offer
compactness, simplicity and a unified framework for computing a wide
range of fluctuation phenomena, allowing techniques and ideas from one
area to be more easily applied to another.

A technique that in principle shares many similarities with the VIE
method is the so-called discrete-dipole approximation
(DDA)~\cite{Purcell73}, which models objects as finite arrays of
polarizable dipoles whose response and interactions due to incident
electromagnetic fields can be obtained via the solution of a
corresponding integral equation~\cite{Yurkin2007}.  DDA has been
recently employed and suggested as an efficient approach for computing
radiative heat transfer~\cite{Edalatpour2014} as well as
fluorescence~\cite{Yurkin2007,LeRu08} from arbitrary geometries, but
unfortunately suffers from a number of important
limitations. Technically, DDA belongs to the general class of volume
integral equations traditionally solved numerically via the method of
weighted residuals~\cite{Finlayson_book} (or method of moments as it
is conventionally known when applied to computational
electromagnetics~\cite{chew01}), by which integral equations are
converted into a solvable and finite set of linear systems of
equations. Specifically, system unknowns (fields or equivalent
currents) are approximated by expanding them in a finite set of basis
functions, often determined by discretizations of objects into meshes
or grids, and then forcing the resulting semi-discrete equations to be
equal in a weak sense, i.e. by integrating them against a set of
testing functions~\cite{Markkanen2012}. The actual choice and
combination of basis and testing functions gives rise to a plethora of
practical variants~\cite{Markkanen2012}.

DDA can be considered to be a particular implementation of the VIE
method known as a collocation method~\cite{Harrington_book}, involving
constant or dipole basis functions and Dirac-delta distributions for
testing, with solutions forced to be accurate only at a finite set of
points (known as point matching)~\cite{Harrington_book}. However, it
is now known that methods of weighted residuals are only guaranteed to
\textit{converge in norm} under special circumstances, the lack of
which can lead to numerous convergence and efficiency
issues~\cite{Buffa_book}. Specifically, basis functions must span the
function space of the unknowns and testing functions must span the
dual space of the range of the corresponding VIE
operator~\cite{vanBeurden2007,vanBeurden2008}. DDA respects neither of
these, and as a consequence its applicability is largely limited to
situations involving light scattering in structures with small index
contrasts and weakly polarizable media~\cite{Yurkin2007}, beyond which
it can lead to a number of severe convergence and accuracy
problems~\cite{edalatpour2015convergence}.  (Note that DDA also makes
a number of other approximations that break down in geometries
involving wavelength-scale objects, cf. Eq.~14 in
\citeasnoun{Yurkin2007}.) In contrast, our FVC formulation is based on
a recently developed VIE framework (dubbed JM-VIE) that is numerically
solved by means of a Galerkin method of
moments~\cite{Polimeridis2014}. JM-VIE exploits basis and testing
functions spanning the function space of internal volume
currents~\cite{Polimeridis2014}, the stability and superior
convergence of which have been demonstrated in geometries involving
highly inhomogeneous objects and large dielectric
contrasts~\cite{Polimeridis2014}. While the associated JM-VIE matrix
elements involve complicated, expensive, and highly singular
volume--volume integrals of homogeneous Green's functions integrated
against pairs of basis functions, these were recently shown to reduce
to surface--surface integrals over smoother kernels that can be
readily handled using specialized integration techniques originally
developed for SIE methods~\cite{DEMCEM,DIRECTFN}.

In the following sections, we derive our FVC formulation of
fluctuating currents and demonstrate that it can be employed to study
a wide class of electromagnetic fluctuation effects in general
geometries, with no uncontrolled approximations except for the finite
discretization (basis). We begin in \secref{FVC-deriv} with a brief
review of the VIE formulation of electromagnetic scattering, followed
by derivations of formulas involving power and momentum transfer, as
well as far-field radiation patterns from radiating objects. The final
\textit{boxed} expressions are described via traces of products of VIE
and current--current correlation matrices which encode the spatial and
spectral characteristics of the fluctuating sources. In \secref{FTC},
we show that important algebraic properties of the associated VIE and
correlation matrices allow efficient evaluation of the trace
expressions; specifically, a number of the VIE matrices admit low-rank
approximations, enabling us to exploit sophisticated and fast
iterative techniques for their evaluation. Finally, in \secref{apps}
the FVC framework is validated against known results and also applied
to obtain predictions in new geometries that currently lie outside the
scope of state-of-the-art techniques, such as objects subject to
spatially varying temperatures and dielectric properties.

\section{FVC formulation}
\label{sec:FVC-deriv}

In this section, we begin by reviewing the VIE method of EM scattering
and apply it to derive an FVC formulation of fluctuation-induced
phenomena in inhomogeneous media. Our approach relies on the JM-VIE
formulation and associated Galerkin method of moments presented
in~\citeasnoun{Polimeridis2014}, also briefly discussed. As noted
above, a strategy based on SIE formulations is unavailable for
modeling inhomogeneous objects since finding the radiation of a point
source (the Green's function) in inhomogeneous media is nearly
impossible with only surface unknowns~\cite{Oijala2014}. Matters are
further complicated for fluctuation phenomena involving power or
momentum transfer, in which case inhomogeneities in the properties of
the fluctuating sources (e.g. spatial variations throughout the bodies
due to temperature or dielectric changes) must also be accurately
accounted for. Starting with the recently developed power
formulas~\cite{Polimeridis2015}, we derive compact trace expressions
for the power and momentum transfer and far-field radiation pattern of
complicated objects with inhomogeneous properties. Finally, we
elaborate on special algebraic properties of the associated VIE and
correlation matrices that allow fast computations of the matrix-trace
formulas, making large and complicated calculations tractable.

\begin{figure}[t!]
\begin{center}
\includegraphics[scale=0.4]{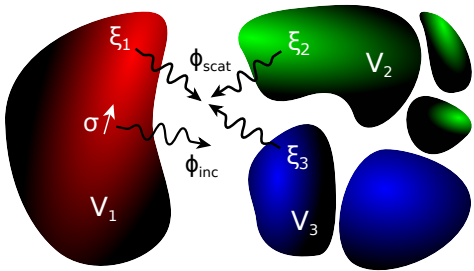}
\caption{Schematic of a many-body geometry in which fluctuating
  current sources give rise to radiation as well as flux and momentum
  transfer between the bodies. Also illustrated are the incident field
  $\phi_\mathrm{inc}$ due to a single dipole source $\sigma$ within a
  body $V_1$ along with the induced polarization--currents $\xi$
  throughout $V_1$ and two nearby bodies, $V_2$ and $V_3$, resulting
  in scattered fields $\phi_\mathrm{scat}$. The characteristics of the
  dipole sources $\sigma$ (fluctuation statistics) and the
  permittivities of the bodies $\chi$ (material properties) both vary
  within each object.}
\label{fig:schem}
\end{center}
\end{figure}

\subsection{Volume integral equations}
\label{sec:VIE}

The derivations of VIEs often rely on the volume equivalence
principle, which shares many similarities with---but is significantly
simpler and more easily derived than---the more well-known surface
equivalence principle~\cite{Stratton39,Chen89,Harrington89}. Consider
the system of arbitrarily shaped, inhomogeneous bodies described by
the relative permittivity $\epsilon$ and permeability $\mu$ functions,
depicted schematically in \figref{schem}. Let $\phi$ and $\s$ denote
6-component electromagnetic fields and volume currents,
\begin{equation*}
  \phi = \begin{pmatrix} \vec{E} \\ \vec{H} \end{pmatrix}, \quad \s
= \begin{pmatrix} \vec{J} \\ \vec{M} \end{pmatrix}.
\end{equation*}
and consider the scattering problem involving incident fields
$\phi_{\rm inc}$ due to $\s$ (in the absence of bodies) and scattered
fields $\phi_{\rm scat}$ due to reflections and scattering from
objects and sources. Defining the 6-component volume currents
\begin{equation}
\label{eq:EJdef}
\xi = \begin{pmatrix} \vec{J}_b \\ \vec{M}_b \end{pmatrix}
= -i \omega \chi \phi
\end{equation}
associated with bound polarization $\vec{J}_b$ and magnetization
$\vec{M}_b$ currents inside the objects, described by the $6\times 6$
susceptibility tensor $\chi$ (which for convenience also includes the
permittivity and permeability of the ambient medium), it follows that
the scattered field can be be written as a convolution of $\xi$ with
the homogeneous Green's function of the ambient
medium~\cite{Chew_book}. (Note that there is no assumption on $\chi$,
which can describe both anisotropic and/or chiral media, changing only
the form of the homogeneous Green's function~\cite{Staelin94}.) In
particular, the unknown scattered fields can be shown to be related to
the free and bound currents, respectively, via convolutions $(\star)$
with the $6\times 6$ homogeneous Green's tensor of the ambient medium
(typically free space) $\Gamma(\ro,\rs)=\Gamma(\ro-\rs,\vec{0})$,
written explicitly in \citeasnoun{RodriguezReid12:FSC}. This is the
core idea behind the volume equivalence principle, which we review
below.

We begin by writing the total field $\phi = \Gamma \star (\s + \xi)$
via the volume equivalence principle~\cite{Chew_book} in terms of the
incident $\phi_{\rm inc} = \Gamma \star \s$ and scattered $\phi_{\rm
  scat} = \Gamma \star \xi$ fields, or more explicitly:
\begin{equation}
\label{eq:Efield}
\begin{split}
\phi(\ro)&= \int d^3\rs \, \Gamma(\ro,\rs) \left[ \sigma(\rs) + \xi(\rs) \right]
\end{split}
\end{equation}
where it is clear that all of the scattering information (including
material inhomogeneities) is ``encoded'' in the convolution of the
homogeneous Green's function with the polarization/magnetization
current. Multiplying both sides of \eqref{Efield} with $-i \omega
\chi$ and using the definition of $\xi$ in \eqref{EJdef}, one arrives
at the following VIE for the induced currents $\xi$:
\begin{equation}\label{eq:VIEop}
  \xi + i \omega \chi (\Gamma \star \xi)= -i \omega \chi (\Gamma \star \sigma),
\end{equation}
which can be solved to obtain $\xi$ from the incident sources
$\s$. This is the so-called JM-VIE formulation of electromagnetic
scattering in which the unknowns are induced currents rather than
fields or field densities. Compared to other formulations based on
field unknowns, JM-VIE exhibits superior performance in terms of
accuracy and convergence, especially for objects with high refractive
index~\cite{Markkanen2012b, Polimeridis2014}.

The operator equation above is customarily solved by reducing it to an
approximate, finite-dimensional linear system. Let $\{b_\alpha\}$ be
some convenient set of $N$ vector-valued basis functions. We can then
approximate our unknowns $\xi$ (and, for convenience below, the source
currents $\sigma$) in this basis:
\begin{equation}
\xi(\ro) \approx \sum\limits_{\alpha=1}^N x_\alpha b_\alpha(\ro), \quad
\sigma(\ro) \approx \sum\limits_{\alpha=1}^N s_\alpha b_\alpha(\ro).
\end{equation}
There are two main categories of basis functions that are used in the
numerical solution of the JM-VIE above, known as \textit{spectral} and
\textit{MoM sub-domain} bases. A spectral basis consists of
non-localized Fourier-like basis functions whereas MoM sub-domain
bases are localized functions obtained by discretizing objects into
meshes or grids of volumetric elements, e.g. cubes, tetrahedra, and
hexahedra~\cite{Kolundzija_book}, and defining functions by low-order
polynomials with local support in one or a few elements. In this work,
we resort to the second category and exploit piecewise constant basis
functions defined in cubes, due to the flexibility they offer for
modeling geometries of arbitrary shape~\cite{Polimeridis2014}. We note
however that the proposed framework and the resulting matrix-trace
formulas can also be evaluated using spectral bases as well.

Finally, the semi-discrete equation is ``tested'' with another set of
functions (called testing functions) to produce a linear system. In
the Galerkin approach, the set of testing functions is the same with
the one of the basis functions. The resulting Galerkin JM-VIE linear
system reads
\begin{equation}\label{eq:VIE_system}
\mat{W}^{-1}\,\matvec{x} = (\mat{V} - \mat{W}^{-1}) \,\matvec{s},
\end{equation}
where
\begin{equation}\label{eq:WVMatrices}
\begin{split}
\mat{W}^{-1}_{\alpha,\beta} &= \langle b_\alpha,b_\beta + i \omega \chi (\Gamma \star b_\beta) \rangle\\
\mat{V}_{\alpha,\beta} &= \langle b_\alpha , b_\beta \rangle
\end{split}
\end{equation}
and $\alpha,\beta=1:N$. Also, $\langle , \rangle$ denotes the standard
inner product of functions $\langle \phi , \psi \rangle = \int
\phi^\ast \psi$, with the $\ast$ superscripts denoting the conjugate
transpose (adjoint) operation. Without loss of generality, we can
choose the basis functions to satisfy an orthogonality relation, so
that $\langle b_\alpha , b_\beta \rangle = \delta_{\alpha \beta}$. In
this case the matrix $\mat{V}$ (often called Gram matrix) is equal to
the identity matrix, i.e., $\mat{V}\equiv\mat{I}$, and it follows that
\begin{equation}
\matvec{x} + \matvec{s} = \mat{W}\mat{V} \,\matvec{s} = \mat{W} \,\matvec{s}.
\end{equation}
Note that our simplifying assumption of orthogonal basis
functions can be easily relaxed, leading to slightly modified $W \to W
V$ and $C \to C V$ matrices (below).

The numerical evaluation of Galerkin inner products in
\eqref{WVMatrices} involves multidimensional integrals over the
support of both basis and testing functions. This integration can be
quite cumbersome due to singularities (when the support of the basis
and the testing functions overlap) and the highly dimensional aspect
of the problem. However, previous work\cite{Polimeridis2013b}
demonstrated that these challenging volumetric integrals can be
reduced to surface integrals (of lower singularity), allowing us to
benefit from decades of work dedicated to the accurate and efficient
evaluation of the associated surface integrals. Here, we make use of
the free-software DEMCEM\cite{DEMCEM} and DIRECTFN\cite{DIRECTFN},
which leverage the techniques described in
\citesasnoun{Polimeridis2013b, Polimeridis2013c} . Furthermore, MoM
JM-VIE formulations with local basis/testing functions typically
result in very large linear systems, which can be solved with
iterative algorithms for non-symmetric dense systems. In each
iteration, the associated matrix-vector products take ${\cal{O}} (N^2
)$ time. Moreover, it is practically impossible to explicitly store
the (dense) matrix $\mat{W}^{-1}$ requiring ${\cal{O}} (N^2 )$
memory. In fact, there are now well-established, fast algorithms to
reduce the costs of such integral equation
solvers~\cite{Philips1997,Jarvenpaa2013,Polimeridis2014}. However, the
ability to exploit fast solvers in fluctuation EM problems is not a
priori guaranteed since as we show below the final formulas involve
complicated traces of products of JM-VIE and related matrices. In
\secref{FTC}, we describe a fast procedure for the computation of the
proposed matrix-trace, which relies on a straightforward and easily
implemented FFT-based fast algorithm presented
in~\citeasnoun{Polimeridis2014} that scales as ${\cal{O}}(N \log{N} )$
for each matrix-vector product and requires ${\cal{O}}(N )$ memory.

Before concluding this section, we introduce some additional
definitions and notation. In particular, further below we exploit the
so-called Green matrix $\mat{G}$, defined as
\begin{equation}\label{GMatrix}
\mat{G}_{\alpha,\beta} = \langle b_\alpha, \Gamma \star b_\beta \rangle,
\end{equation}
which involves interactions among basis functions mediated by the
Green's function. For $n$ objects, the associated matrices and vectors
can be conveniently written as:
\begin{equation}
\mat{G}\rightarrow
\begin{pmatrix}
\mat{G}^{11}  &  \mat{G}^{12} & \cdots & \mat{G}^{1n}   \\[0.5em]
\mat{G}^{21}  &  \mat{G}^{22} & \ldots & \vdots               \\[0.5em]
\vdots        &  \vdots       & \ddots &  \vdots    \\[0.5em]
\mat{G}^{n1}  &  \mat{G}^{n2} & \cdots & \mat{G}^{nn}
\end{pmatrix};\quad
{\xi}\rightarrow
\begin{pmatrix}
\mat{ \xi}^{1}    \\[0.5em]
\mat{\xi}^{2}    \\[0.5em]
 \vdots   \\[0.5em]
 \mat{ \xi}^{n}
\end{pmatrix}
\end{equation}
where the superscripts denote blocks associated with the various
objects, with diagonal components corresponding to self-interactions
and off-diagonal blocks involving interactions between different
objects. Finally, we define the projection,
\begin{equation}
\mat{P}^p_{\alpha,\beta} =
\begin{cases}
1, &\mbox{if } \alpha = \beta =  p\\
0, &\mbox{otherwise},
\end{cases}
\end{equation}
which selects specific blocks of vectors $\hat{x}^p = P^p x$ or
diagonal blocks of matrices $\hat{A}^p = P^p A P^p$ corresponding to
object $p$.

\subsection{Power transfer}
\label{sec:PTF}

We now derive a compact matrix-trace formula for the computation of
the ensemble-averaged flux into body ${\cal{B}}_p$ (or equivalently
the absorbed power) due to fluctuating current sources in body
${\cal{B}}_q$, integrated over all possible positions and
orientations. The first step consists of the evaluation of the flux
from ${\cal{B}}_p$ due to a single dipole source $\s$ immersed in
${\cal{B}}_q$, which we denote as $\Phi^{q\rightarrow
  p}_{\sigma}$. Direct application of Poynting's theorem implies that
the flux on the objects is given by:~\cite{Jackson_book}
\begin{equation}
  \Phi^{q\rightarrow p}_{\sigma} = \frac{1}{2} \Re{}
  \int_{{\cal{B}}_p} d^3\ro \, \xi^\ast \cdot \phi
\end{equation}
which amounts to the work done by the total field on the polarization
currents in ${\cal{B}}_q$. Expressing the induced currents and fields in the basis of JM-VIE currents and using the relation $\phi = \Gamma \ast (\xi + \sigma)$
yields the following discrete approximation (see \citeasnoun{Polimeridis2015} for a complete analysis):
\begin{equation}\label{eq:Pabs}
\begin{split}
\Phi^{q\rightarrow p}_{\sigma}  &=  \frac{1}{2}\Re{} \matvec{x}^{p\ast}{\phi}^p
= \frac{1}{2}\Re{} \matvec{x}^{\ast} \mat{P}^p {\phi}\\
&=  \frac{1}{2}\Re{} ( \matvec{x} + \hat{\matvec{s}}^q)^{\ast} \mat{P}^p \mat{G} ( \matvec{x} + \hat{\matvec{s}}^q)\\
&= \frac{1}{2} ( \matvec{x} + \hat{\matvec{s}}^q)^{\ast} \sym{(\mat{P}^p \mat{G})} ( \matvec{x} + \hat{\matvec{x}}^q)\\
&= \frac{1}{2} \left( \mat{W} \mat{P}^q \matvec{s}\right)^{\ast} \sym{(\mat{P}^p \mat{G})} \left( \mat{W}\mat{P}^q \matvec{s}\right)\\
&= \frac{1}{2} \trace{ ( \matvec{s} \matvec{s}^\ast) (\mat{W}\mat{P}^q)^\ast \sym{(\mat{P}^p \mat{G} )} (\mat{W}\mat{P}^q) }
\end{split}
\end{equation}
where $\sym{\mat{G}} = \frac{G+G^*}{2}$ denotes the Hermitian part of
$\mat{G}$. It is then straightforward to obtain the ensemble-averaged
flux $\Phi^{q\rightarrow p} \equiv \langle \Phi^{q\rightarrow
  p}_{\sigma} \rangle$, which yields:
\begin{equation}\label{eq:PTF}
\begin{split}
\Phi^{q\rightarrow p} &= \frac{1}{2} \trace{ \langle \matvec{s}
  \matvec{s}^\ast\rangle (\mat{W}\mat{P}^q)^\ast \sym{(\mat{P}^p
    \mat{G})} (\mat{W}\mat{P}^q) }\\ &= \frac{1}{2} \trace{\mat{P}^q
  \mat{C} \mat{P}^q \mat{W}^\ast \sym{(\mat{P}^p \mat{G})}
  \mat{W}}
\end{split}
\end{equation}
where $\mat{C}=\langle \matvec{s} \matvec{s}^\ast\rangle$ is a
current--current correlation matrix that captures a statistical,
ensemble average over sources, described in more detail in
\secref{Cmat}. Defining the matrix $\hat{C}^q = P^q C P^q$, which is
simply a projection of the correlation matrix unto the space of basis
functions in $q$, we find that the ensemble-averaged flux is given by:
\begin{equation}
\boxed{
  \Phi^{q\rightarrow p} = \frac{1}{2} \trace{ \hat{\mat{C}}^q
    \mat{W}^\ast \sym{(\mat{P}^p \mat{G})} \mat{W}}.
}
\end{equation}

\subsection{Momentum transfer}
\label{sec:MTF}




In addition to carrying energy, the radiation emitted by fluctuating
sources also carries linear and angular momentum, which can also be
described using similar expressions.  The starting point consists of
the evaluation of the force (or torque) imparted on an object
${\cal{B}}_p$ due to a single dipole source immersed in
${\cal{B}}_q$. Although electromagnetic forces are often computed via
surface-integrals of the Maxwell stress tensor, it is also possible
and in our case more convenient to express the force as a volume
integral by considering the Lorentz force acting on the internal
currents $\xi$ induced on ${\cal{B}}_p$~\cite{Krueger2012}. In
particular, the force on the object is given by:
\begin{align}
\label{eq:force}
\vec{F}^{q\to p}_{\sigma} &= \frac{1}{2\omega}\text{Im } \int_{{\cal{B}}_p}
  d^3\ro\, \xi^* \cdot \nabla \phi
\end{align}
where $\nabla$ denotes the usual partial derivative with respect to
infinitesimal displacements. The derivation of the above expression
follows from application of the time-average 
Lorentz force $d\vb F = \frac{1}{2}\text{Re }(\rho^* \vb E + \vb J^* \times \vb B) d^3\ro$ on the electric charge and current densities ($\rho, \vb J$)
in an infinitesimal volume element $d^3\ro$, together with a similar expression for the force on the magnetic sources. Integrating over the volume of the body and employing Stokes' theorem along with Maxwell's equations
immediately yields \eqref{force}. In a similar fashion, the torque
about some origin $\ro_0$ can be obtained by integrating the
differential torque $d\boldsymbol{\tau} = (\ro-\ro_0) \times d\vb F$
on a volume element.

Expressing the induced currents and fields in the basis of JM-VIE
currents and following a similar procedure as that of \secref{PTF},
one finds that the ensemble-averaged force on the object can be
written in the compact and convenient form:
\begin{equation}
\boxed{
  \vec{F}^{q \to p} = \frac{1}{2\omega} \Tr 
  \Big[ \hat{C}^q W^* \asym \left(P^p \GF \right) W\Big],
}
\label{eq:transfer}
\end{equation}
where in this case and in contrast to power transfer, the relevant
quantity is the matrix representation $\GF$ of the gradient of the
Green's function operator $G$, whose matrix elements
$\GF_{\alpha,\beta} = \langle b_\alpha, \nabla \Gamma \star
b_\beta\rangle$. Also, $\asym{\mat{G}} = \frac{G-G^*}{2}$ denotes the
skew-Hermitian part of $\mat{G}$. The torque on the object can be
obtained similarly by computing angular derivatives of $G$. It turns
out that the calculation of these matrix elements requires evaluating
multidimensional integrals whose singularities are more severe than
those of $G$.  A key distinction between fluctuation-induced transfers
of power and momentum is that, in the latter case, one finds nonzero
fluctuation-induced forces and torques between bodies even at thermal
equilibrium and even at zero temperature; these are just the usual
equilibrium Casimir forces.~\cite{ReidRo12:review} \Eqref{transfer},
which computes only the non-equilibrium contribution to the force,
must generally be augmented by these equilibrium contributions to
yield the total force. Connections between \eqref{transfer} and
expressions for equilibrium forces, along with techniques for
evaluating the above-mentioned integrals and results of VIE
computations of non-equilibrium Casimir forces and torques are
addressed in subsequent work~\cite{Reid2015}.

\subsection{Far-field radiation intensity}
\label{sec:DIR}

In addition to power and momentum transfer, another useful quantity is
the far-field radiation intensity of our system, which can also be
expressed as a simple trace formula. The result which follows
trivially from \eqref{PTF}, is that the ensemble-averaged flux
radiated by an isolated body ${\cal{B}}_q$ to the background medium is
given by:
\begin{equation}
\label{eq:PTF_0}
\Phi^{q\rightarrow 0}  
= -\frac{1}{2} \trace{ \mat{C} \mat{W}^\ast \sym{ \mat{G}} \mat{W}}
\end{equation}
where the minus sign corresponds to the direction of the power flux
and stems from Poynting's theorem. However, in addition to the overall
radiation, it is also useful to obtain the radiation intensity over
specific directions, or equivalently the power radiated per solid
angle. The angle-resolved radiation intensity $U_\sigma^{q\rightarrow
  0}$ from a single source $\s$ immersed in ${\cal{B}}_q$ can be
obtained by expressing the radiation field at infinity
$\vec{E}_\infty$ (where only far field contributions remain) in terms
of the free and bound current sources, as follows:
\begin{equation}
U^{q\rightarrow0}_{\sigma} = \frac{k^2 Z}{2(4\pi)^2} |Q
\vec{e}_\infty(\ro)|^2 = \frac{k^2 Z}{2(4\pi)^2} |Q
\left[\Gamma_\infty^{\rm E} \star (\sigma + \xi) \right]|^2
\end{equation}
where $k$ is the wavenumber and $Z=\sqrt{\mu_0/\epsilon_0}$ is the
wave impedance, both in vacuum.  Also, $\Gamma_\infty^{\rm
  E}(\ro,\rs)$ is the $3\times 6$ Green's tensor of the ambient medium
which maps currents to far-field electric fields, and $Q$ is a
$3\times 3$ transformation tensor that maps vectors from Cartesian to
spherical coordinates and projects their radial component to
zero~\cite{Balanis_book}. Given the solution of the VIE scattering
problem and following the same procedure described above, it is
straightforward to write the radiation intensity as a matrix-trace
formula of the form:
\begin{equation}
\begin{split}
U^{q\rightarrow0}_{\sigma} &= \frac{k^2 Z}{2(4\pi)^2} (\matvec{s}+\matvec{x})^\ast \mat{G}_{\infty}^{\rm E \ast} \mat{G}_{\infty}^{\rm E}(\matvec{s}+\matvec{x})\\
&= \frac{k^2 Z}{2(4\pi)^2} \trace{(\matvec{s} \matvec{s}^\ast)(\mat{W})^\ast (\mat{G}_{\infty}^{\rm E\ast}\mat{G}_{\infty}^{\rm E}) (\mat{W})}
\end{split}
\end{equation}
where the matrix $\mat{G}_{\infty}^{\rm E}$ is the discretized form of
the operator $Q \Gamma_\infty^{\rm E}$, obtained in a similar fashion
as $\mat{G}$. Ensemble averaging over all sources, we find that the
final formula for the angle-resolved radiation intensity
$U^{q\rightarrow0} \equiv \langle U^{q\rightarrow0}_{\sigma} \rangle$
is given by:
\begin{equation}
  \label{eq:Uformula}
\boxed{ U^{q\rightarrow0} = \frac{k^2 Z}{2(4\pi)^2}
  \trace{\mat{C}\,\mat{W}^\ast (\mat{G}_{\infty}^{\rm
      E\ast}\mat{G}_{\infty}^{\rm E}) \mat{W}}.  }
\end{equation}
\Eqref{Uformula} can be integrated over all solid angles $\Omega$ to
yield the total radiation rate $\Phi^{q\rightarrow0} = \int d\Omega \,
U^{q\rightarrow0}(\Omega)$, which as expected agrees with results
obtained by direct application of \eqref{PTF_0}, as discussed in
\secref{FTC}.


\subsection{Current--current correlation matrices}
\label{sec:Cmat}

The formulas above are very general in that they apply to many
different kinds of fluctuation processes, the physical properties and
origins of which are embedded in the correlation matrices $C = \langle
s s^*\rangle$, involving ensemble averages over all sources $\s$ and
polarizations throughout the bodies. In particular, the matrix
elements of the correlation matrices describe interactions among basis
functions and are given by:
\begin{equation}
\begin{split}
  \mat{C}_{\alpha,\beta} = \langle s_\alpha s_\beta^\ast \rangle =
  \int \int d^3\ro d^3\rs \, b_{\alpha}^\ast(\ro) \langle \sigma(\ro) \sigma^\ast(\rs)
  \rangle b_{\beta}(\rs) 
 \end{split}
\end{equation}
which follows trivially from the orthogonality property of our basis
functions and the fact that $\s(\vec{x}) = \sum_\alpha s_\alpha
b_\alpha(\vec{x})$. Although in general the calculation of each matrix
element involves volume--volume integrals against pairs of basis
functions, current fluctuations are temporally and spatially
uncorrelated in local media~\cite{Agarwal74,Eckhardt84,Matloob97} and
are described by:
\begin{equation}
  \langle \sigma_i(\ro,\omega) \sigma_j^\ast(\rs,\omega) \rangle =
  \mathcal{J}_{ij}(\ro,\omega)\delta(\ro-\rs)
\label{eq:JJ}
\end{equation}
where the subscripts denote polarization degrees of freedom and
$\mathcal{J} \geq 0$ is a position-dependent spectral tensor whose
form depends on the physical origins of the fluctuations. It follows
that $C$ is Hermitian and positive-semidefinite and thus admits a
Cholesky factorization $\matvec{C} =
\mat{L}_{\mat{C}}\mat{L}_{\mat{C}}^\ast$, which we exploit in
\secref{FTC} to demonstrate that our radiation, power, and momentum
formulas are susceptible to fast-trace calculations.

When the sources of fluctuations involve only quantum and thermal
vibrations (heat), the correlation function $\mathcal{J}$ is
determined by thermodynamic considerations such as the well-known
FDT~\cite{Lifshitz80,Eckhardt84}, relating current fluctuations to
dissipation in materials. Without loss of generality, the spectral
function is given by:~\cite{Lifshitz80}
\begin{equation}
  \mathcal{J}_{ij}(\ro,\omega) = \frac{4}{\pi} \Im \chi_{ij}(\ro,\omega)
  \Theta(\ro,\omega),
\label{eq:FDT}
\end{equation}
where the $\Im \chi$ tensor describes losses in the medium and
$\Theta(\vec{x},\omega) = \hbar \omega / (e^{\hbar \omega/k_{\rm B}
  T(\vec{x})} - 1)$ is the Planck distribution, or the average energy
of an oscillator having local temperature $T(\vec{x})$. \Eqref{FDT} in
conjunction with the power transfer and radiation formulas above are
exploited below to evaluate thermal radiation and heat transfer
between inhomogeneous bodies with spatially varying temperature and
dielectric properties, and also in an upcoming paper that focuses on
non-equilibrium Casimir forces~\cite{Reid2015}.

In situations involving active media driven by external pumps, the
properties of the fluctuating currents and hence $\mathcal{J}$ depend
on the details of the input drive along with the physical emission
mechanisms. For a broad range of processes, the spectral function can
be written in the simple form:
\begin{equation}
\label{eq:fluo}
\mathcal{J}_{ij}(\ro,\omega) = \chi_{\rm inc}(\ro) \chi_{\rm
  emm, ij}(\vec{x},\omega),
\end{equation}
where $\chi_{\rm inc}$ describes the response of the medium due to the
pump and $\chi_{\rm emm}$ describes the emission spectrum of the
excited medium, which depends on the distribution of active molecules
in the medium and on complicated electronic transitions mediated by
the pump as well as quantum/thermal processes~\cite{LeRu08}. In the
particular example of one-photon fluorescence from a medium (with high
quantum yield) excited by incident light, the pump spectrum is
proportional to the locally absorbed power and hence can be computed
by direct application of the VIE power formulas. Such a relationship
in conjunction with \eqref{Uformula} is exploited below to compute the
fluorescence spectrum of an irradiated sphere. A similar dependence on
the local field intensity arises in the case of Raman scattering,
except that $\chi_{\rm inc}$ is proportional to the Raman
polarizability tensor rather than the susceptibility of the
medium~\cite{LeRu08}. In the case of spontaneous emission from a gain
medium, the emission spectrum is determined by spatially dependent
effective permittivity and temperature profiles determined by the
driven steady-state atomic populations of the medium, both of which
can be obtained by application of steady-state ab-initio laser theory
(SALT)~\cite{Henry86,Matloob97}. Similar descriptions apply in more
complicated systems, including fluorophores with low quantum yields or
active media subject to highly nonlinear (e.g. two-photon) processes.



\section{Fast Trace Computations}
\label{sec:FTC}

The matrix-trace formulas derived in the previous sections require
products of inverses of the JM-VIE matrix $\mat{W}$ with dense
matrices $\sym{(\mat{P}^p \mat{G})} $, $\asym{(\mat{P}^p \nabla
  \mat{G})}$, and $\mat{G}_{\infty}^{\rm E\ast}\mat{G}_{\infty}^{\rm
  E}$. As mentioned above, due to their large size and correspondingly
severe CPU and memory limitations, it is practically impossible to
form explicitly either the Green matrix or its inverse. There are
however fast FFT-based procedures for evaluating matrix-vector
products of the JM-VIE system matrix and the Green
matrix~\cite{Polimeridis2014}. Here we describe a framework based on
iterative methods for the fast computation of the associated trace
formulas above.

We begin with the matrix-trace formula $\Phi^{q\rightarrow p}$ in the
presence of $n$ bodies (including ${\cal{B}}_p$ and ${\cal{B}}_q$),
which after some algebraic manipulations can be written as follows
(ignoring pre-factors):
\begin{equation}\label{eq:fastPTF}
\begin{split}
\Phi^{q\rightarrow p}
&=
\trace{ \mat{C}^{qq} \mat{W}^{pq\ast}(\sym{\mat{G}^{pp}})\mat{W}^{pq} }\\
&+ \sum\limits_{\substack{m=1 \\ m\neq p}}^n\trace{ \mat{C}^{qq}   \sym{\left( \mat{W}^{pq\ast}\mat{G}^{pm} \mat{W}^{mq} \right)}  } \\
&= S^{q\rightarrow p} + C^{q\rightarrow p}
\end{split}
\end{equation}
where $\mat{C}^{qq}$ is the $qq$ block of the matrix $\mat{C}$. Due to
the different characteristics of $S^{q\rightarrow p}$ and
$C^{q\rightarrow p}$, we need to address them separately. As discussed
in \secref{Cmat}, the matrix $\matvec{C}^{qq}$ can be assumed to be
Hermitian and positive semidefinite, hence it admits a Cholesky
factorization, $\matvec{C}^{qq} =
\mat{L}_{\mat{C}^{qq}}\mat{L}_{\mat{C}^{qq}}^\ast$. In addition,
$\sym{\mat{G}^{pp}}$ is a Hermitian, negative semidefinite
matrix\cite{Rodriguez2013} and it also admits a low-rank approximation
since it is associated with the smooth, imaginary part of the Green's
functions. Hence, it can be approximated to any desired accuracy by a
truncated singular value decomposition (SVD) factorization,
$\sym{\mat{G}^{pp}} \approx
-\mat{U^{pp}}\mat{S^{pp}}\mat{U^{pp}}^\ast$, where $\mat{S^{pp}} \in
\mathbb{C}^{r\times r}$, with $r \ll N$. The norm of the error in the
aforementioned truncation is bounded by the norm of the vector of
discarded singular values. The classical SVD algorithm requires the
complete matrix, hence we resort here to a class of modern randomized
matrix approximation techniques, and more specifically to the
randomized SVD method (rSVD)~\cite{Halko2011,Hochman2014}. rSVD is
effective for matrices with fast drop of the singular values and it
requires only a fast matrix-vector procedure, which we have developed
as described above. The matrix with the singular values can be further
decomposed so that $\mat{S^{pp}} =
\mat{L_{S^{pp}}}\mat{L_{S^{pp}}^\ast}$. Finally, it follows that the
self-term in \eqref{fastPTF} can be written as the square of a
Frobenius norm,
\begin{equation}
\begin{split}
S^{q\rightarrow p}  &= -\trace{ \mat{L}_{\mat{C}^{qq}} \mat{L}_{\mat{C}^{qq}}^\ast (\mat{W}^{pq\ast} \mat{U^{pp}}) \mat{L_{S^{pp}}}\mat{L_{S^{pp}}^\ast} (\mat{U^{pp\ast}} \mat{W}^{pq}) }\\
&=  - \| \mat{L}_{\mat{C}^{qq}}^\ast (\mat{W}^{pq\ast} \mat{U^{pp}})\mat{L_{S^{pp}}} \|_{\rm F}^2.
\end{split}
\end{equation}
For the most time consuming part of the norm, we need to solve the
adjoint JM-VIE system $r$ times (for each of the leading singular
vectors of $\sym{\mat{G}^{pp}}$). Note however that we can solve for
each vector of $\mat{U^{pp}}$ independently and thus the entire
procedure is embarrassingly parallelizable. Also,
$\mat{L}_{\mat{C}^{qq}}^\ast$ and $\mat{L_{S^{pp}}}$ are either sparse
or diagonal, while $\mat{W}^{pq\ast} \mat{U^{pp}}$ is a
``tall-and-skinny'' matrix (the number of columns is much smaller than
the number of rows) and hence the matrix product appearing in the norm
can be computed efficiently.

The trace formula for $C^{q\rightarrow p}$ is not symmetrical and
therefore cannot be reduced to a norm. In this case, one can exploit
the fact that $\mat{G}^{pm}$ admits a low-rank approximation due to
the smoothing properties of the Green's function for disjoint
objects. The final dimensions of the low-rank approximation of
$C^{q\rightarrow p}$ (for a prescribed accuracy) depend on the
electric distance between objects $p$ and $m$ \cite{Chai2013}, i.e.,
$\mat{G}^{pm} \approx \mat{U^{pm}}\mat{S^{pm}}\mat{V^{pm}}^\ast$,
where $\mat{S^{pm}} \in \mathbb{C}^{l\times l}$, with $l \ll N$. The
final formula for $C^{q\rightarrow p}$ after the Cholesky
factorization of the singular values matrix ($\mat{S^{pm}}$) is given
by
\begin{equation}\label{eq:C_qp}
\begin{split}
C^{q\rightarrow p} =\Re{} \sum\limits_{\substack{m=1 \\ m\neq p}}^n
\trace{\mat{X_\mat{U^{pm}}} \mat{X_\mat{V^{pm}}^\ast} }
\end{split}
\end{equation}
where
\begin{equation*}
\begin{split}
\mat{X_\mat{U^{pm}}} &= \mat{L}_{\mat{C}^{qq}}^\ast  (\mat{W}^{pq\ast} \mat{U^{pm}}) \mat{L}_{\mat{S^{pm}}}\\
\mat{X_\mat{V^{pm}}} &= \mat{L}_{\mat{C}^{qq}}^\ast  ( \mat{W}^{mq\ast} \mat{V^{pm}} ) \mat{L}_{\mat{S^{pm}}}.
\end{split}
\end{equation*}
Both $\mat{X_\mat{U^{pm}}}$ and $\mat{X_\mat{V^{pm}}} $ are
``tall-and-skinny'', and we can not compute the trace by forming
explicitly their product, due to memory limitations. Alternatively, we
can use the standard vectorization of a matrix ${\rm vec} ()$, which
converts the matrix into a column vector, together with the identity,
$\trace{\mat{X}\mat{Y}^\ast} = {\rm vec}(\mat{X})^T \cdot
\overline{{\rm vec}(\mat{Y})}$, and write \eqref{C_qp} in the
following computationally friendly form:
\begin{equation}
\begin{split}
C^{q\rightarrow p}
= \Re{} \sum\limits_{\substack{m=1 \\ m\neq p}}^n
 {\rm vec}(\mat{X_\mat{U^{pm}}} )^T \cdot  \overline{{\rm vec}(\mat{X_\mat{V^{pm}}}) }.
\end{split}
\end{equation}
The overall computational complexity for the evaluation of
$C^{q\rightarrow p}$ consists of a single run of the Randomized-SVD
for a non-symmetric matrix\cite{Halko2011}, and $2\times l$ solves of
the adjoint JM-VIE system. In the case of the matrix-trace formulas
for the force and the torque, the procedure is similar with the one
described above. The only difference stems from the replacement of
$\mat{G}$ with $\GF$ and $\sym{}$ with $\asym{}$.

Finally, the case of far-field radiation is somewhat
simpler. According to \eqref{Uformula}, we just need to solve $2$
times the adjoint JM-VIE system, since $\mat{G}_{\infty}^{\rm
  E\ast}\in\mathbb{C}^{N\times 2}$. Hence, the radiation intensity for
a specific direction or solid angle $\Omega$, is given by the
following square of the Frobenius norm:
\begin{equation}\label{eq:U_fro}
\begin{split}
U^{q\rightarrow p}(\Omega)
=  \frac{k^2 Z}{2(4\pi)^2}   \| \mat{L}_{\mat{C}}^\ast (\mat{W}^\ast \mat{G}_{\infty}^{\rm E\ast})  \|_{\rm F}^2.
\end{split}
\end{equation}
This is a very useful formula, especially when directional information
of the radiated power is of interest. In addition, the total radiated
power can be evaluated by integrating \eqref{U_fro} over all solid
angles, as mentioned in \secref{DIR}, which would amount to employing
a numerical integration scheme over the unit sphere (e.g. Lebedev
quadrature~\cite{Lebedev1976}). Alternatively, one could exploit
\eqref{PTF_0} and the associated norm $\| \mat{L}_{\mat{C}}^\ast
(\mat{W}^{\ast} \mat{U})\mat{L_{S}} \|_{\rm F}^2$ to compute the total
radiated power from an isolated body. The latter is expected to be
more efficient for total-radiation computations with prescribed
accuracy, controlled by the SVD factorization of the Green matrix, in
which case the minimum number of JM-VIE solves needed for a prescribed
accuracy is estimated in advance. In contrast, the former approach is
based on adaptive quadrature schemes where the accuracy is controlled
by the comparison of results between different orders of integration,
with no a priori control.


\section{Validation and Applications}
\label{sec:apps}

In this section, we apply the FVC method to obtain new results in
complex geometries. To begin with, we show that the Green matrices
appearing in our trace formulas admit low-rank decompositions (as
discussed in \secref{FTC}) by computing their ranks to within some
tolerance in a representative structure involving two
vacuum-separated, homogeneous cubes. We validate the FVC method by
checking its predictions against known results of thermal radiation
and near-field heat transfer between homogeneous bodies, including
spheres, cubes, and ellipsoids, obtained using a boundary-element
implementation of our recent FSC
formulation~\cite{RodriguezReid12:FSC}. We show that when subject to
temperature gradients or continuously varying permittivities, complex
bodies can exhibit highly modified thermal radiation and heat transfer
spectra, leading to directional emission at selective
wavelengths. Finally, we demonstrate that the same formalism can be
exploited to study luminescence from excited media by computing the
fluorescence spectrum of a sphere irradiated by monochromatic incident
light. We show that the impact of the resulting inhomogeneous current
fluctuations cannot be easily obtained by exploiting simple
homogenization or effective-medium approximations. For convenience and
simplicity, we consider dielectric media with no material dispersion
(constant $\Re \epsilon \approx 12$ and large dissipation $\Im
\epsilon\approx 1$), though our approach is general in that it can
readily handle other kinds of materials such as metals with $\Re
\epsilon < 0$ and even gain media.

\subsection{Low-rank approximations}

Low-rank approximations of the associated (free-space) Green matrices
are instrumental to the practical and efficient evaluation of our
trace formulas. In this section, we present some representative
results obtained from computing the ranks of both $\sym{\mat{G}^{pp}}$
and $\mat{G}^{pm}$, to within some tolerance, for the particular
problem of two vacuum-separated, homogeneous cubes of edge-length
$L=2R$ and separated by a surface--surface distance $d$, shown
schematically in \figref{heat}.

\begin{figure}[t!]
\begin{center}
  \includegraphics[scale=0.42]{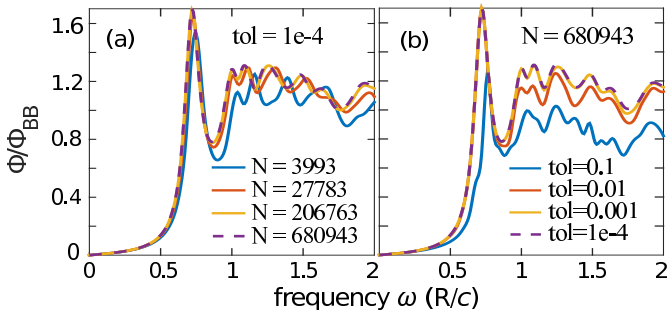}
  \caption{Flux spectrum $\Phi(\omega)$ of a cube of edge-length $2R$
    held at temperature $T$, normalized by the corresponding
    black-body spectrum $\Phi_\mathrm{BB}(\omega) = \frac{A}{4\pi^2}
    (\omega/c)^2\Theta(\omega,T)$, for different (a) discretization
    mesh densities and (b) rSVD truncation tolerances.}
  \label{fig:convergence}
\end{center}
\end{figure}

\begin{table}[t!]
\renewcommand{\arraystretch}{1.0}
\caption{Ranks of $\sym{\mat{G}^{11}}$ for various frequencies ($\omega \frac{R}{c}$) and tolerances (tol) in truncated SVD. The ranks correspond to the case of a cube of edge-length $2R$. In addition, results for a sphere of radius $R$ are included in brackets.} \label{symG_svd} \centering
\footnotesize
\begin{tabular}{c | c c c c c c}
\hline\hline
\diaghead(4,-3){\theadfont Diag Head}%
{$\omega \frac{R}{c}$ \\}{ tol \\}
& $1e^{-1}$     & $1e^{-2}$        & $1e^{-3}$         & $1e^{-4}$      & $1e^{-5}$        & $1e^{-6}$         \\[0.4em] \hline \\
0.01 & 4 (4)         & 4 (4)            & 4 (4)             & 4 (4)          & 7 (7)            & 12 (12)         \\[0.3em]
0.1  & 4 (4)         & 4 (4)            & 7 (7)             & 12 (12)        & 12 (12)          & 14 (12)         \\[0.3em]
1.0  & 12 (7)        & 14 (12)          & 24 (24)           & 40 (24)        & 40 (40)          & 60 (40)         \\[0.3em]
2.0  & 18 (12)       & 37 (24)          & 51 (40)           & 65 (60)        & 84 (60)          & 109 (84)       \\[0.3em]
\hline\hline
\end{tabular}
\end{table}

\begin{table}[t!]
\renewcommand{\arraystretch}{1.0}
\caption{Ranks of $\mat{G}^{12}$ for various distances ($d$) and tolerances (tol) in truncated SVD. The ranks correspond to the case of two cubes of edge length $L=2R$ and frequency $\omega \frac{R}{c} = 1$. Each cube is discretized into $N=40^3$ voxels, resulting in $3N$ total degrees of freedom, i.e., $\# \rm{DOFS} = 3N$.} \label{AscaH_svd} \centering
\footnotesize
\begin{tabular}{c | c c c c c c}
\hline\hline
\diaghead(5,-3){\theadfont Diag Head}%
{$d/L$ \\}{ tol \\}
         & $1e^{-1}$     & $1e^{-2}$     & $1e^{-3}$     & $1e^{-4}$    & $1e^{-5}$     & $1e^{-6}$         \\[0.4em] \hline \\
0.001& 4075            & 4853              & 5253            & 6352            & 7240            & 8481        \\[0.3em]
0.01  & 992              & 2611             & 3934            & 4800            & 5832            & 6894        \\[0.3em]
0.1    & 50                & 196               & 447              & 804              & 1268            & 1849         \\[0.3em]
1.0    & 6                  & 14                 & 27                & 42                & 66                & 89         \\[0.3em]
10.0  & 4                  & 7                   & 9                  & 14                & 19                & 23       \\[0.3em]
\hline\hline
\end{tabular}
\end{table}

Table I shows the singular values of $\sym{\mat{G}^{11}}$,
corresponding to one of the two cubes, as a function of the normalized
frequency $\omega R/c$ and tolerance $\rm tol$; that is, we obtain the
singular values that produce SVD factorizations bounded in norm by the
tolerance $\rm tol$, also known as a truncated SVD. Since the
associated matrix is very large and our trace formulations can be cast
in terms of fast matrix--vector products, our calculations exploit the
rSVD method recently developed for big-data
problems~\cite{Halko2011}. (Note that results for the second cube,
involving $\sym{\mat{G}^{22}}$, would be identical since both cubes
have equal sizes and number of unknowns.) Our results reveal at least
two important features: First, the ranks scale linearly with $\omega$
at large frequencies, and sub-linearly (roughly constant) at small
frequencies. Additional numerical experiments (not shown) confirm that
the effect of mesh density on the ranks is negligible, yet another
manifestation of the favourable convergence properties of the JM-VIE
formulation~\cite{Polimeridis2014}. This also suggests a strategy for
obtaining the finite rank of $\sym{\mat{G}^{pp}}$ with prescribed
accuracy: we begin by computing the rank of the operator for a
prescribed accuracy by using a coarse mesh and then run a fixed-rank
rSVD algorithm with finer mesh. Finally, \figref{convergence}
illustrates the rate of convergence of the radiation spectrum
$\Phi(\omega)$ from an isolated cube at a fixed temperature $T$ with
respect to different (a) discretization mesh densities and (b)
truncation tolerance, normalized to the spectrum of a corresponding
black body $\Phi_\mathrm{BB}(\omega) = \frac{A}{4\pi^2}
(\omega/c)^2\Theta(\omega,T)$, where $A$ denotes the surface area of
the cube.

The situation changes in the case of the ``coupling'' Green matrix
$\mat{G}^{12}$, which encodes interactions between objects. Table II
shows the significant singular values associated with the coupling
matrix of the same cube--cube geometry at a fixed frequency $\omega$
and for various separations $d$, obtained by leveraging the rSVD
technique. As expected, the singular values increase as $d$ decreases,
a consequence of the power-law drop-off of the Green's function with
separation in the near field. It follows that the computation
complexity of the trace formulas increases as the two bodies come
close together. (Note that, as described in \secref{FTC}, our trace
formulas for power and momentum transfer require us to solve two VIE
systems for every corresponding eigenvector, but fortunately each
system can be solved independently and the overall process is
embarrassingly parallelizable.) Nevertheless, we find that $G^{12}$
remains very low rank even for relatively close separations $d/L
\approx 0.1$, below which constraints on the resolution make the FVC
approach less practical. However, it is precisely at such small
separations that approximate methods such as the proximity
approximation become accurate~\cite{Sasihithlu11}.



\subsection{Thermal radiation and heat transfer}
\label{sec:thermal}

\begin{figure}[t!]
\begin{center}
\includegraphics[scale=0.8]{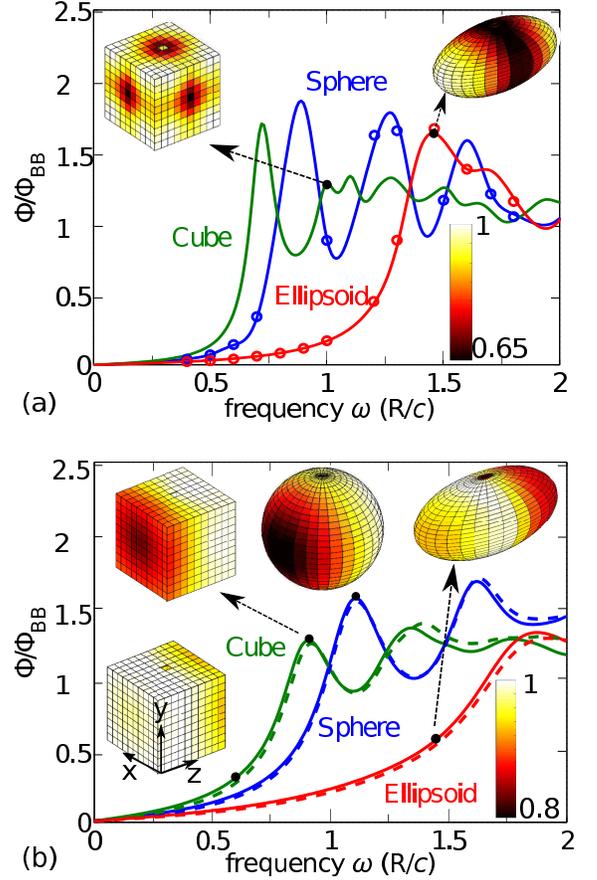}
\caption{Flux spectrum $\Phi(\omega)$ normalized by the corresponding
  black-body spectrum $\Phi_\mathrm{BB}(\omega) = \frac{A}{4\pi^2}
  (\omega/c)^2\Theta(\omega,T)$ of different bodies of surface area
  $A$ held at temperature $T=1000$~K, including a sphere of radius $R$
  (blue lines), cube of edge-length $2R$ (green line), and ellipsoid
  of long semi-axis $R$ and short semi-axis $\frac{R}{2}$ (red
  line). The objects have either (a) uniform permittivities
  $\epsilon=12+i$ or (b) spatially varying $\epsilon(z) =
  \epsilon_{-R} + (\epsilon_R-\epsilon_{-R}) \frac{|z+R|}{2R}$, with
  $\epsilon_R=12+i$ and $\epsilon_{-R}=2+i$. For comparison, we also
  plot the radiation spectrum $\Phi_{\rm eff}$ (dashed lines) of
  corresponding bodies with homogeneous effective permittivities
  $\epsilon_{\rm eff} = 7+i$. The insets depict the angular
  distribution of far-field radiation $U(\Omega)$, normalized by the
  maximum intensity over all directions $\max_\Omega U$, at selected
  frequencies.}
\label{fig:radiation}
\end{center}
\end{figure}



We begin by validating our FVC approach by checking its predictions of
thermal radiation from homogeneous bodies against results obtained
using our recently developed FSC
formulation~\cite{RodriguezReid12:FSC,RodriguezReid12:long}, which is
well-suited for handling piece-wise constant structures and
fluctuations statistics. \Figref{radiation}(a) shows the flux spectra
$\Phi(\omega)$ of multiple objects (of uniform temperature $T$ and
permittivity $\epsilon=12+i$, including a sphere of radius $R$ (blue
line), a cube of edge-length $2R$ (green line), and an prolate
ellipsoid of long semi-axis $R$ and short semi-axes $\frac{R}{2}$ (red
line).  Note that in each case $\Phi(\omega)$ is normalized to the
corresponding flux from a black body. As shown, there is excellent
agreement between the FVC (solid lines) and FSC (circles) predictions,
both of which illustrate the expected radiation enhancement at
geometric resonances.

The FVC method can also handle more complex structures, including
inhomogeneous bodies with spatially varying permittivities. In
particular, \figref{radiation}(b) shows $\Phi(\omega)$ for the same
geometries of \figref{radiation}(a) but for objects with linearly
varying permittivity profiles $\epsilon(z) = \epsilon_{-R} +
(\epsilon_R-\epsilon_{-R}) \frac{|z+R|}{2R}$, with $\epsilon_{-R} =
2+i$ and $\epsilon_R=12+i$ (solid lines) and axes chosen to lie at the
geometric center of each object. Compared to the spectrum of the
homogeneous bodies of \figref{radiation}(a), one finds that the
resonances are shifted to larger frequencies and their peak amplitudes
are significantly smaller, a consequence of the decreased effective
permittivity of each object. For comparison, we also show $\Phi_{\rm
  eff}(\omega)$ (dashed lines) from corresponding homogeneous objects
with effective permittivities,
\begin{equation}
  \epsilon_{\rm eff} = \frac{1}{V}\int_V d^3\vec{x}\, \epsilon(\vec{x}),
\end{equation}
corresponding to uniform $\epsilon_{\rm eff}=7+i$. Our calculations
reveal that in the illustrated frequency range and for our choice of
dielectric profiles, the homogeneous approximation is qualitatively
accurate to within $10\%$. On the other hand, employing
\eqref{Uformula} to compute the angular radiation patterns at selected
frequencies, shown as insets in \figref{radiation}, reveals
significant changes, e.g. significantly larger directional emission,
that cannot be captured by the effective-medium approximation. In
particular, the radiation patterns of the inhomogeneous objects break
$\hat{z}$ mirror symmetry. For example, the flux from the cube at
$\omega\approx 0.65R/c$ is slightly larger in the $-\hat{z}$ than in
the $+\hat{z}$ direction, a situation that is reversed at larger
$\omega\approx 0.9R/c$ (see insets). Generally, the transition
frequency of the favored radiation direction depends on the geometry;
for instance, even at a frequency as high as $\omega \approx 1.5R/c$,
the ellipsoid continues to radiate more along the $-\hat{z}$
direction.

\begin{figure}[t!]
\begin{center}
\includegraphics[scale=0.8]{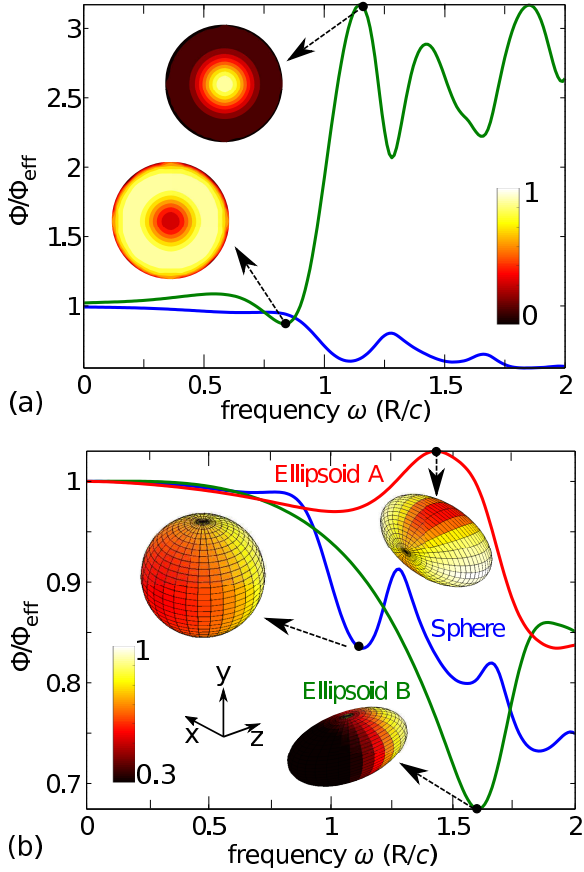}
\caption{Flux spectrum $\Phi(\omega)$ of various bodies normalized by
  the corresponding predictions of a simple approximation
  $\Phi_\mathrm{eff}$, defined in~\eqref{Teff}, including (a) sphere
  of radius $R$ and radially varying temperature profile $T(r) =
  T_0+(T_R-T_0) \frac{r}{R}$ for both $T_0=0,T_R=1000$~K (blue line)
  and $T_0=1000,T_R=0$ (green line), and (b) sphere of radius $R$
  (blue line) or ellipsoids with short semi-axes $\frac{R}{2}$ and
  long semi-axis $R$ along the $\hat{z}$ (green line) or $\hat{x}$
  (red line) directions, subject to vertically varying temperature
  profiles $T(z) = T_{-L} + (T_L-T_{-L}) \frac{|z+L/2|}{L}$, where $L$
  denotes the $z$-dimension of the corresponding body.  In all cases,
  objects have uniform permittivity $\epsilon=12+i$ and are subject to
  temperature gradients $T_{-L}=0$ and $T_{L}=1000$~K. The insets in
  (a) show the local density of states along a cross-section of the
  sphere at different frequencies while those in (b) show the angular
  distribution of far-field radiation $U(\Omega)$ normalized by
  $\max_\Omega U$.}
\label{fig:radiationT}
\end{center}
\end{figure}

More pronounced changes arise when objects are subject to spatial
temperature gradients. \Figref{radiationT} shows $\Phi(\omega)$ from
homogeneous ($\epsilon=12+i$) ellipsoids subject to either (a)
radially varying $T(r) = T_0 + (T_R-T_0) \frac{r}{R}$ or (b)
$z$-varying temperature profiles (see caption). In both cases, $\Phi$
is normalized by the flux $\Phi_\mathrm{eff}$ obtained from a naive
approximation in which the temperature variations are removed in favor
of a uniform effective temperature $T_\mathrm{eff}$ determined by a
simple average of the Planck distribution over the volume $V$ of the
bodies,
\begin{equation}
  \Theta(\omega,T_\mathrm{eff}) = \frac{1}{V} \int_V
  d^3\vec{x}\,\Theta(\omega,T(\vec{x})).
\label{eq:Teff}
\end{equation}
Such a simple approximation obviates the need for exact calculations
that explicitly incorporate inhomogeneities, but is clearly inadequate
for wavelength-scale objects. Specifically, \figref{radiationT}(a)
shows $\Phi(\omega)$ from spheres with radially varying temperatures,
illustrating that beyond the sub-wavelength regime $\omega \ll R/c$
and depending on the choice of $T_0$ and $T_R$, $\Phi$ can be many
times larger or smaller than that predicted by \eqref{Teff}. The
failure of this naive approximation is especially apparent near
resonances, where the coupling of fluctuating sources (dipoles) to
far-field radiation (the local density of states) is highly
position-dependent. The insets of \figref{radiationT}(a) show
cross-sections of the spatially varying flux contribution from dipoles
in the interior of the sphere at two relatively close frequencies. At
$\omega R/c \approx 1.1$, we find that dipoles closer to the center
can couple more efficiently to far-field radiation than those near the
edges, causing \eqref{Teff} to underestimate the flux by
$\Phi/\Phi_\mathrm{eff} \approx 3$ in the case $T_0=0$, $T_R=1000$~K
(green line) and to overestimate it by $\Phi/\Phi_\mathrm{eff} \approx
0.8$ when $T_0=1000~\mathrm{K}$, $T_R=0$ (blue line). The converse is
true at $\omega R/c \approx 0.85$, in which case their coupling to
radiation is largest at the center and edges of the sphere. Similar
effects arise in situations involving $z$-varying temperature
profiles, explored in \figref{radiationT}(b) for either spheres (blue
line) or ellipsoids with either their long-axes (green line) or
short-axes (red line) aligned with the $\hat{z}$ direction. For
instance, ellipsoids can exhibit highly directional emission (almost a
factor of 3 times larger) along the direction of increasing
temperature.


\begin{figure}[t!]
\begin{center}
\includegraphics[scale=0.45]{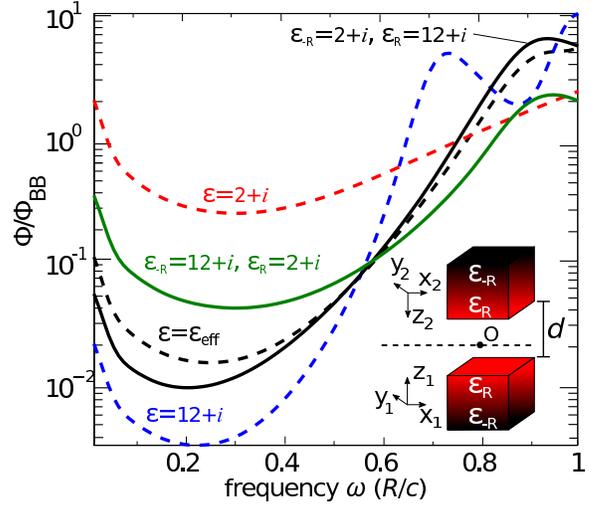}
\caption{Heat-transfer spectrum $\Phi(\omega)$, normalized by the
  corresponding black-body spectrum $\Phi_\mathrm{BB}(\omega) =
  \frac{A}{4\pi^2} (\omega/c)^2\Theta(\omega,T)$, between two cubes of
  edge-length $2R$ and temperature $T=1000$~K separated by
  surface--surface distance $d=R$. The cubes are assumed to have
  either uniform permittivities $\epsilon=2+i$ (red dashed line),
  $\epsilon=\epsilon_\mathrm{eff}=7+i$ (black dashed line), or
  $\epsilon=12+i$ (blue dashed line), or vertically varying
  permittivities $\epsilon(z_i) = \epsilon_{-R} +
  (\epsilon_R-\epsilon_{-R}) \frac{|z_i+R|}{2R}$ defined with respect
  to the local axis $\vec{x}_{1,2}$ at the center of each cube (shown
  on the inset), chosen so that the system has mirror symmetry about
  the $x$--$y$ plane intersecting the origin $O$.  The gradients are
  either increasing (black solid line) or decreasing (green solid
  line) toward or away from the center, corresponding to the choice of
  $\epsilon_{R,-R}=\{12+i,2+i\}$ or $\epsilon_R \leftrightarrow
  \epsilon_{-R}$, respectively.}
\label{fig:heat}
\end{center}
\end{figure}


In addition to far-field radiation, the FVC method can be employed to
obtain radiative transfer between objects. \Figref{heat} shows the
heat-transfer spectrum $\Phi(\omega)$ (computed via \eqref{transfer})
normalized by $\Phi_\mathrm{BB}(\omega)$ (same as above), between two
vacuum-separated cubes of edge-length $2R$ and surface--surface
separation $d=R$, of either uniform (dashed lines) or vertically
varying (solid lines) permittivities. We consider dielectric profiles
of the form $\epsilon(z_i) = \epsilon_{-R} +
(\epsilon_R-\epsilon_{-R}) \frac{|z_i+R|}{2R}$ defined with respect to
the local axis located at the center of each cube $\vec{x}_{1,2}$,
chosen so that the entire system has mirror symmetry about the origin
(see inset). We consider two different profiles,
$\epsilon_{-R,R}=\{2+i,12+i\}$ (black line) or $\epsilon_R
\leftrightarrow \epsilon_{-R}$ (green line), corresponding to
increasing gradients toward or away from the origin. For comparison,
we also plot the transfer between cubes of uniform permittivities
$\epsilon=2+i$ (red dashed line), $\epsilon=12+i$ (green dashed line),
and $\epsilon=\epsilon_{\rm eff} = \frac{1}{V}\int_V d^3\vec{x} \,
\epsilon(z)$, corresponding to the minimum, maximum, or average of the
spatially varying permittivities, respectively. As shown, depending on
the wavelength regime (near versus far field) inhomogeneities can have
a different effect on the heat transer. For instance, at low $\omega
R/c \ll 1$ where near-field effects prevail, homogeneous bodies with
smaller dielectric constants tend to transfer more heat---the same
dependence is observed for planar objects separated by vacuum, where
the near-field contribution $\sim (\frac{\Im
  \epsilon}{|\epsilon+1|^2})^2$~\cite{BasuZhang09}. Not surprisingly,
because nearby regions tend to contribute more than far-away regions,
one observes that despite having the same average permittivities
$\epsilon_{\rm eff}$ (dashed blue line), the transfer is sensitive to
the local dielectric variation, exhibiting larger enhancement in the
case where the permittivity is increasing toward (green solid line)
rather than away (black solid line) from the origin. At larger $\omega
R/c \gtrsim 0.5$ where far-field effects begin to dominate, one
observes the opposite behavior, in which case the largest transfer is
obtained for decreasing permittivities toward the origin. Essentially,
as illustrated in \figref{radiation}(b), at sufficiently large
wavelengths, bodies with dielectric gradients tend to radiate along
the direction of increasing permittivity.

\subsection{Fluorescence}

\begin{figure}[t!]
\begin{center}
\includegraphics[scale=0.48]{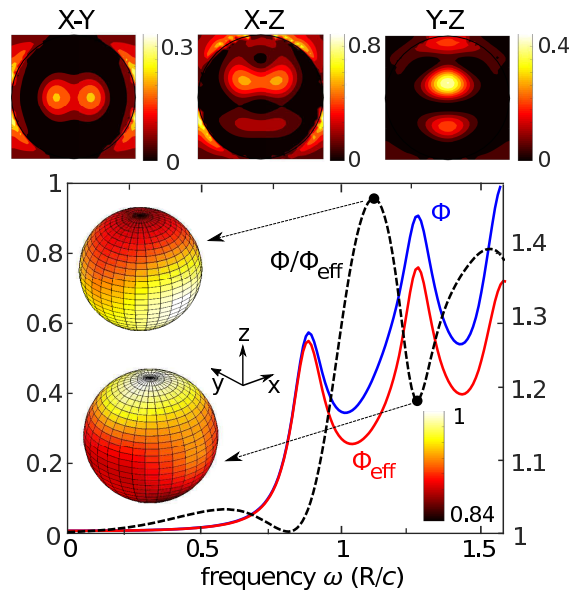}
\caption{Far-field fluorescence spectrum $\Phi(\omega)$ (in arbitrary
  units) of a homogeneous and non-dispersive dielectric sphere of
  radius $R$ and permittivity $\epsilon=12+i$ excited by an
  $\hat{x}$-polarized planewave propagating along the $\hat{z}$
  direction with frequency $\omega_\mathrm{inc} R/c = 1.58$. The
  absorbed power $\chi_{\rm inc}(\vec{x})$ inside the sphere, obtained
  by solving a single scattering problem as described
  in~\citeasnoun{Polimeridis2014}, is shown in the top contour plots
  along three sphere cross-sections.  $\Phi$ is computed exactly (blue
  line) or via a homogeneous approximation $\Phi_{\rm eff}$ in which
  the absorbed power is taken to be uniformly distributed inside the
  sphere and given by $\chi_{\rm eff} = \int_V d^3\vec{x} \chi_{\rm
    inc}(\vec{x})$ (red line). The ratio of the two is plotted as the
  black dashed line on the right axis. The insets depict the angular
  distribution of fluorescence emission, normalized by the maximum
  intensity over all directions, at selected
  frequencies.}\label{fig:fluorescence}
\end{center}
\end{figure}

We now consider application of the FVC formulas to the calculation of
fluorescence. A typical fluorescence setup consists of an incident
wave impinging on a fluorescent body, leading to the absorption and
subsequent re-emission of light by molecules inside the
body.~\cite{LeRu08} Both of these effects are captured by the
current--current correlation matrix described in \secref{Cmat}, which
encodes the spectral properties of the fluctuations. In the particular
problem of one-photon fluorescence induced by an incident
monochromatic wave at a given frequency $\omega_\mathrm{inc}$, the
spectral function $\mathcal{J}(\vec{x},\omega)$ has the form given in
\eqref{fluo}, with the excitation spectrum given by the locally
absorbed power,
\begin{equation}
\chi_{\rm inc}(\vec{x}) \propto \omega_{\rm inc} \Im \chi
|\vec{E}(\vec{x},\omega_{\rm inc})|^2,
\end{equation}
and $\chi_{\rm emm}(\vec{x},\omega)$ denoting the fluorescence
spectrum of the bulk medium, usually a relatively broad Lorentzian
lineshape centered near the material's absorption resonance. (Note
that $\chi_{\rm inc} = 0$ in the absence of a fluorescent medium.) A
well-known approach to enhance fluorescence involves designing bodies
to have strong resonances at $\omega_\mathrm{inc}$, leading to
increased absorption~\cite{LeRu08}. For bodies designed to have
additional resonances within the fluorescence bandwidth, determined by
$\chi_{\rm emm}$, there is an additional source of enhancement arising
from the increased local density of states, or increased coupling of
dipole emitters to far-field radiation. Inhomogeneities arise due to
the fact that $\chi_{\rm inc}$ and the local density of states are
both highly spatially non-uniform near resonances.

\Figref{fluorescence} shows the fluorescence emission $\Phi(\omega)$
from a sphere of radius $R$ and uniform permittivity $\epsilon=12+i$,
irradiated by an $x$-polarized, $z$-traveling incident wave of
frequency $\omega_\mathrm{inc} R/c \approx 1.58$, chosen to coincide
with one of its resonances.  For simplicity, we assume a
non-dispersive and uniformly distributed fluorescent medium with
$\chi_{\rm emm}=1$, although as noted above our formalism can just as
easily handle spatially varying distributions. The first step in
computing the fluorescence emission is to obtain the locally absorbed
power within the sphere $\chi_{\rm inc}(\vec{x})$, which boils down to
the calculation of a single and far simpler scattering problem
exploiting \eqref{Pabs}, as described in
\citeasnoun{Polimeridis2015}. Along with $\Phi$ (blue line),
\figref{fluorescence} shows $\chi_{\rm inc}$ along three different
cross-sections intersecting the center of the sphere (top contour
plots), illustrating the highly non-uniform spatial pattern of current
fluctuations.  Also shown is the spectrum $\Phi_{\rm eff}$ obtained by
application of a homogeneous approximation (red line) where the
absorbed power is averaged over the volume of the sphere to yield a
uniform, effective $\chi_{\rm eff} = \int_V d^3\vec{x}\, \chi_{\rm
  inc}(\vec{x})$, along with the corresponding ratio $\Phi/\Phi_{\rm
  eff}$ (black line). As before, such approximations yield accurate
results in the sub-wavelength regime but break down at larger
frequencies. For instance, at $\omega R/c \approx 1$ we find that
$\Phi/\Phi_{\rm eff} \approx 1.5$. More importantly, the approximation
fails to capture the angular distribution of radiation (insets): both
the direction of largest fluorescence and overall emission pattern
change drastically as the emission frequency increases from $\omega
R/c\approx 1.1$ to $\omega R/c \approx 1.3$.

\section{Concluding remarks}

Our FVC formulation of electromagnetic fluctuations enables accurate
calculations of wide-ranging incandescence (e.g. thermal radiation,
dispersion forces, heat transfer) and luminescence (e.g. spontaneous
emission, fluorescence, Raman scattering) phenomena in arbitrary
geometries. Similar to recently proposed scattering-matrix and
surface-integral equation formulations of radiative heat transfer, the
resulting quantities are obtained via traces of matrices involving
interactions among basis functions; however, because the JM-VIE
``scattering'' unknowns are volume currents rather than propagating
waves or surface currents, the formalism is applicable to a broader
set of problems. For example, as demonstrated here, our approach
captures phenomena associated with the presence material
inhomogeneities, such as spatially varying temperature gradients and
dielectric properties within bodies. In future work, we plan to
exploit the FVC approach to demonstrate predictions of highly
directional radiation from inhomogeneous structures subject to thermal
gradients~\cite{Jin15:thermal}, non-equilibrium Casimir torques on
chiral particles~\cite{Reid2015}, and enhanced directional emission
from parity-time symmetric (gain) media~\cite{Jin15:PT}. Furthermore,
although our calculations focused on geometries involving compact
bodies, the same power and momentum formulas derived above apply to
geometries involving extended bodies, the subject of future work.

\section*{ACKNOWLEDGEMENTS}

This work was supported in part by grants from the Singapore-MIT
programs in Computational Engineering and in Computational and Systems
Biology, from the Skolkovo-MIT initiative in Computational
Mathematics, from the Army Research Office through the Institute for
Soldier Nanotechnologies under Contract No. W911NF-07-D0004, and from
the National Science Foundation under Grant No. DMR-1454836.



\end{document}